\documentclass{iau} 
\usepackage{graphicx}
\usepackage{textcomp}
\usepackage{subcaption}
\usepackage[colorlinks]{hyperref}

\title[IAUS 353: Properties of Barred Galaxies] 
{Properties of barred galaxies in the \\ MaNGA galaxy survey}

\author[Amelia Fraser-McKelvie ]   
{Amelia Fraser-McKelvie$^1$, Michael Merrifield$^{1}$, Alfonso Arag\'{o}n-Salamanca$^{1}$, 
  Karen Masters$^{2}$, \& the MaNGA Survey Team}

\affiliation{$^1$School of Physics \& Astronomy, University of Nottingham, \\ Nottingham NG7 2RD,
U.K. \\ email: {\tt amelia.fraser-mckelvie@nottingham.ac.uk} \\[\affilskip]
$^2$Department of Physics and Astronomy, Haverford College, \\ 370 Lancaster Ave, Haverford, PA 19041, U.S.A. \\
}

\pubyear{2019}
\volume{353}  
\jname{Galactic Dynamics in the Era of Large Surveys}
\editors{J. Shen, M. Valluri, \& J. A. Sellwood  }
\begin{document}

\maketitle

\begin{abstract}
We present the initial results of a census of 684 barred galaxies in the MaNGA galaxy survey. This large sample contains galaxies with a wide range of physical properties, and we attempt to link bar properties to key observables for the whole galaxy. We find the length of the bar, when normalised for galaxy size, is correlated with the distance of the galaxy from the star formation main sequence, with more passive galaxies hosting larger-scale bars. Ionised gas is observed along the bars of low-mass galaxies only, and these galaxies are generally star-forming and host short bars. Higher-mass galaxies do not contain H$\alpha$ emission along their bars, however, but are more likely to host rings or H$\alpha$ at the centre and ends of the bar. Our results suggest that different physical processes are at play in the formation and evolution of bars in low- and high-mass galaxies. 

\end{abstract}

\firstsection 
\section{Introduction}
Galactic bars inhabit the majority of disk galaxies in the present-day Universe. Despite their ubiquity, the community has not yet reached a consensus on the nature of their formation and evolution, nor on the effects of bar dynamics on the host galaxy. Observables such as bar length, strength, and star formation, along with galaxy-wide properties including colour, morphology, gas fraction, and stellar mass have yet to be linked in a way that can reliably describe the effect of a bar on its host galaxy. 
This work attempts to set observational constraints for models of their evolution by performing a census of the bars within 684 galaxies in the MaNGA galaxy survey.  

\section{Sample Selection}
A sample of 684 barred galaxies was selected from the Mapping Nearby Galaxies with APO \cite[(MaNGA; Bundy et al. 2011)]{Bundy11} Galaxy Survey MaNGA Product Release 8 (MPL-8), which contains 6779 unique galaxy observations.
Barred galaxies were selected using Galaxy Zoo 2 classifications \cite[(Willett et al. 2013)]{Willett13}. Using the criteria \texttt{p\_bar\_weighted $>$ 0.5} and \texttt{p\_not\_edgeon $>$ 0.5}, 684 barred galaxies were selected for this analysis.

\section{Results}
\subsection{Bar Length}
\begin{figure}[b]
\begin{center}
 \includegraphics[width=0.99\textwidth]{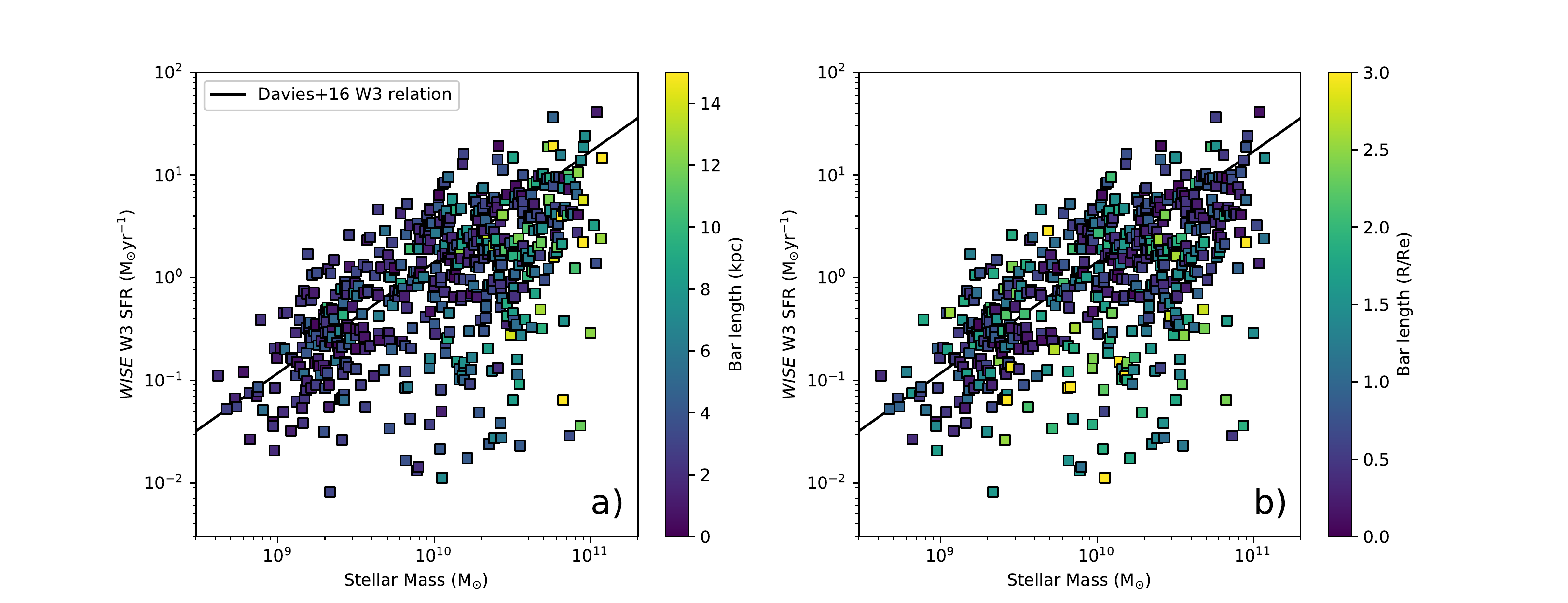} 
 \caption{The star formation main sequence of barred MaNGA galaxies, with main sequence line from \cite[Davies et al. (2016)]{Davies16}. Points are colour coded by bar length in units of kpc (panel a), and $R/R_{\rm{e}}$ (panel b). The scaled bar length of galaxies lying below the main sequence is longer than galaxies still forming stars at the same mass.}
   \label{fig1}
\end{center}
\end{figure}

We determine the length of a bar using the Fourier transform bar analysis code of \cite[Kraljic et al. (2012)]{Kraljic12} on collapsed MaNGA data cube `white-light' images.  
In Fig~\ref{fig1} we present the star formation main sequence (SFMS) of all MaNGA galaxies, with star formation rates (SFRs) derived from 12$\mu$m emission from the \textit{Wide-field Survey Explorer} (\textit{WISE}) AllWISE Source catalogue using the relation of \cite[Cluver at al. (2017)]{Cluver17}. Stellar masses come from the NASA Sloan Atlas catalogue \cite[(NSA; Blanton et al. 2011)]{Blanton11}. From panel a) of Fig~\ref{fig1}, we see that the longest bars are hosted by galaxies with high stellar mass, irrespective of galaxy SFR. The overall size of the host galaxy is taken into account by dividing the bar length by the effective radius of the galaxy. For this we use the $r$-band Petrosian half-light radius from the NSA. From panel b) of Fig~\ref{fig1}, we see that the scaled bar length correlates not with galaxy stellar mass, but with distance from the main sequence. Galaxies further below the main sequence will host bars that are a larger fraction of the overall galaxy size; the more passive the galaxy, the larger-scale bar it hosts. If we assume that bars grow in length with time, this suggests a picture in which bar growth quenches star formation. 

\subsection{Ionised Gas Morphology}
\begin{figure*}
\centering
\begin{subfigure}{0.32\textwidth}
\includegraphics[width=\textwidth]{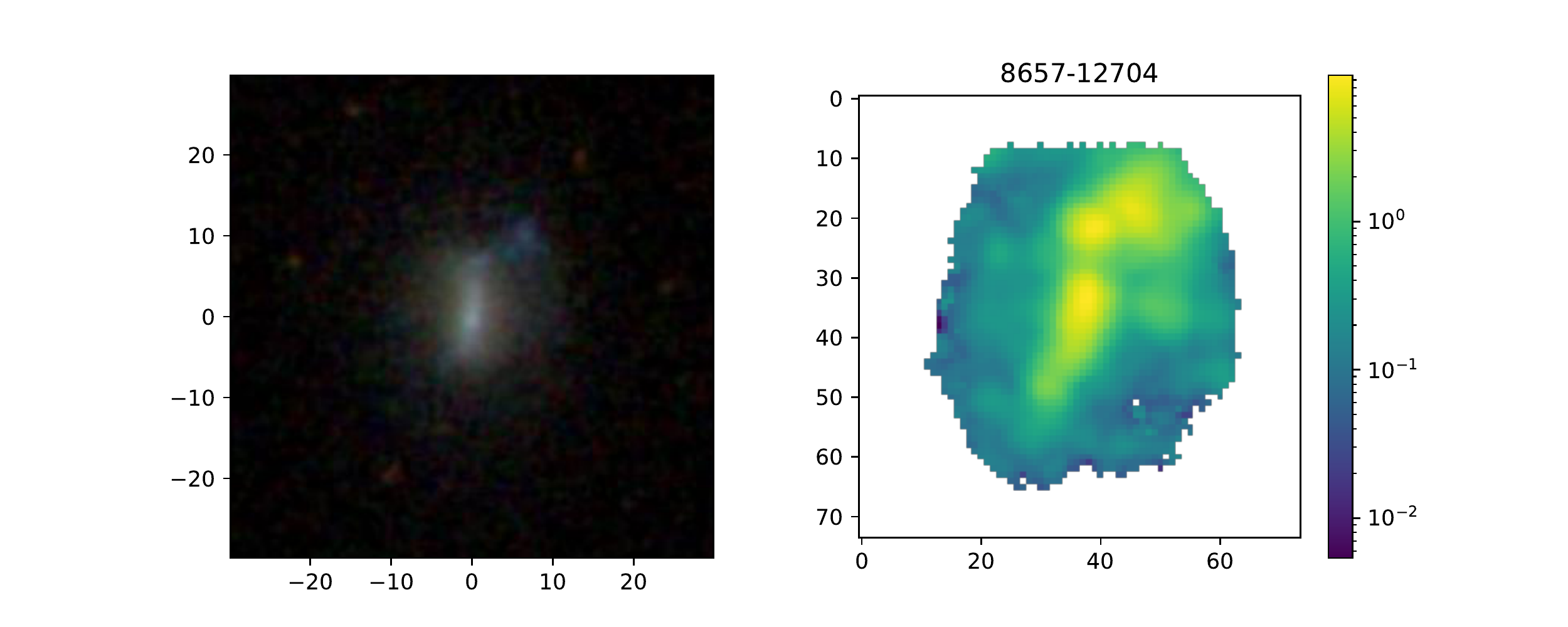} 
\caption{H$\alpha$ present along the bar.}
\end{subfigure}
\begin{subfigure}{0.32\textwidth}
\includegraphics[width=\textwidth]{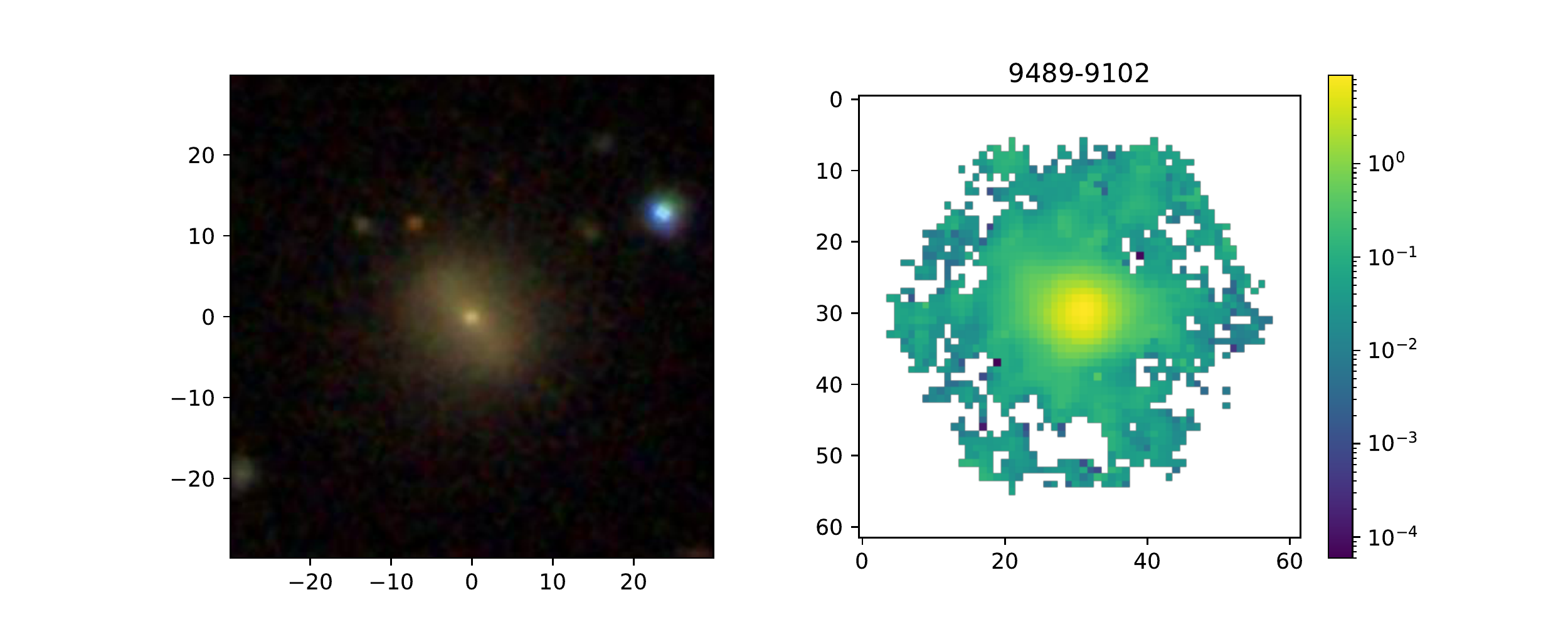} %
\caption{Predominantly central H$\alpha$.}
\end{subfigure}
\begin{subfigure}{0.32\textwidth}
\includegraphics[width=\textwidth]{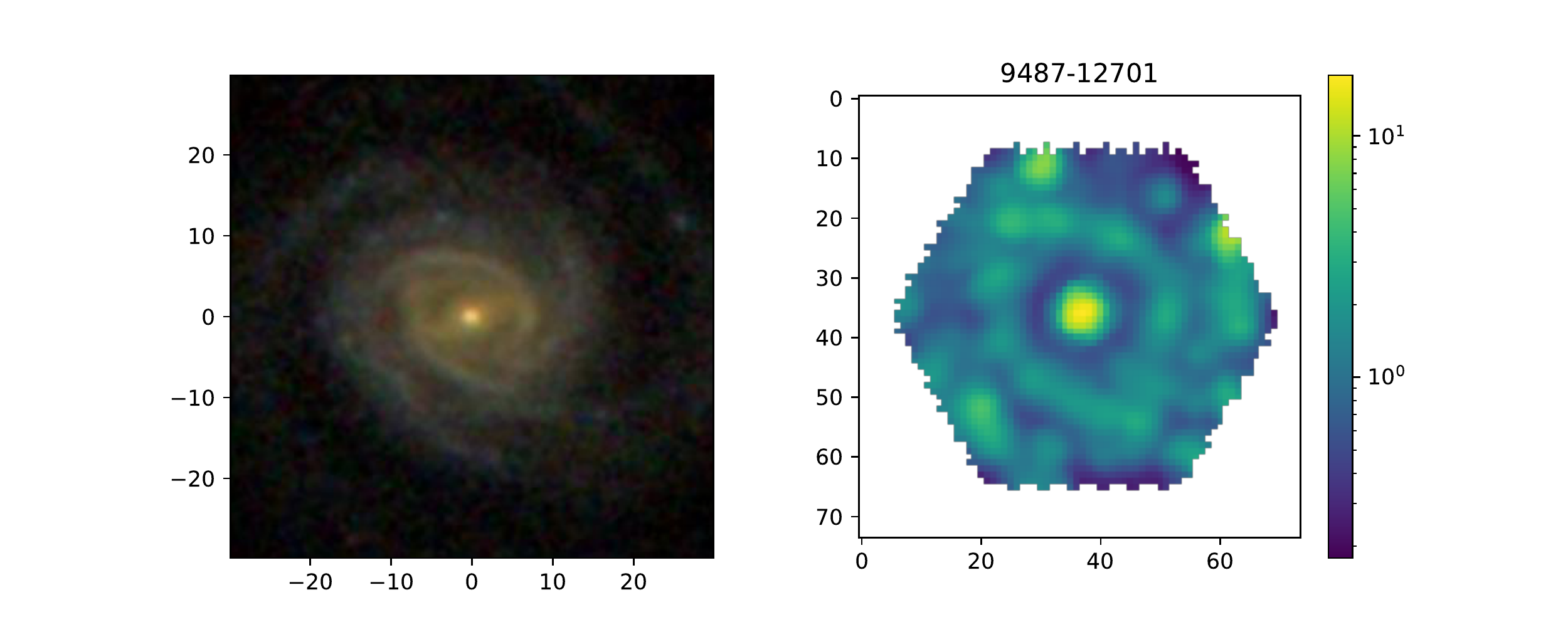} %
\caption{Prominent ring.}
\end{subfigure}
\hfill
\begin{subfigure}{0.33\textwidth}
\includegraphics[width=\textwidth]{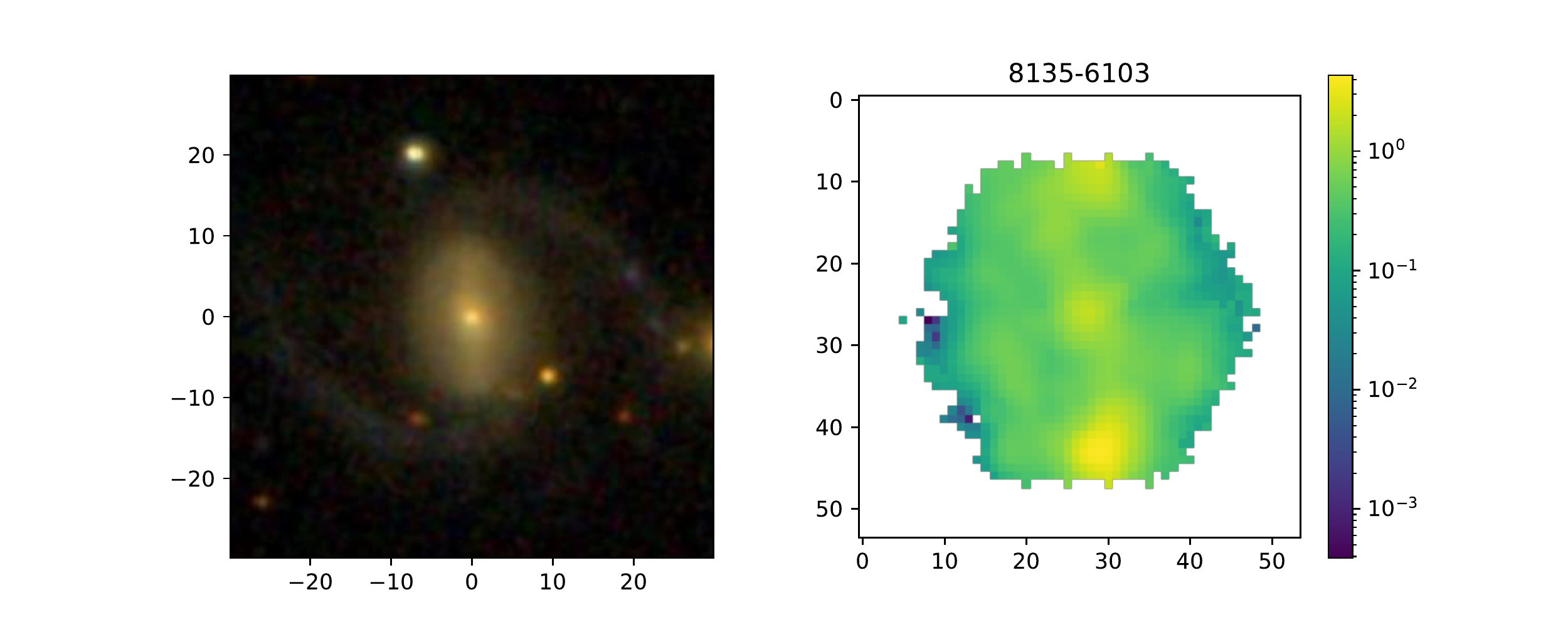} %
\caption{H$\alpha$ at ends of the bar (and sometimes the centre).}
\end{subfigure}
\begin{subfigure}{0.33\textwidth}
\includegraphics[width=\textwidth]{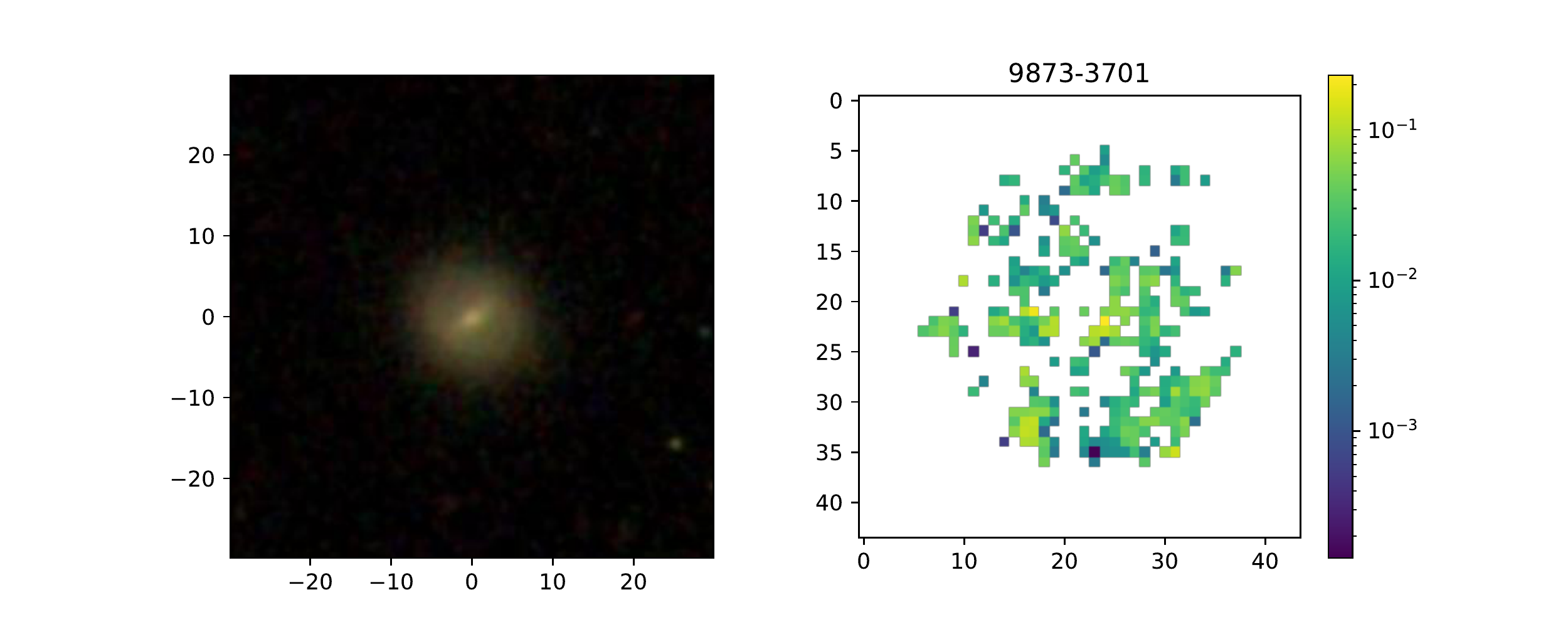} %
\caption{No significant H$\alpha$.\\                       { \color{white} a}   }
\end{subfigure}
\caption{The five H$\alpha$ morphology classifications devised for this work. On the left is a $gri$ image of the galaxy, and on the right, a MaNGA H$\alpha$ map. } %
\label{fig2}
\end{figure*}

From Fig~\ref{fig1} we know the majority of MaNGA barred galaxies are star-forming, so we investigate the spatially-resolved morphology of the ionised gas from the H$\alpha$ emission maps provided by the MaNGA data analysis pipeline (DAP, \cite[Westfall et al. 2019]{Westfall19}). We wish to determine whether the ionised gas morphology correlates with the overall galaxy physical properties, and whether any information on the bar evolution and growth may be drawn from this.
Motivated by works such as \cite[Verley et al. (2007)]{Verley07}, we devised a visual classification scheme, whereby the H$\alpha$ map of a barred galaxy fell into one of five categories. The five categories are listed below, and example maps shown for each classification in Figure~\ref{fig2}.
\begin{itemize}
\item \textbf{H$\alpha$ present along bar.} Extended H$\alpha$ emission was detected across the galaxy, coinciding (or close to) the position of the bar in the optical $gri$ image of the galaxy. If any H$\alpha$ emission was seen along the bar, it was classified into this category.

\item \textbf{Predominantly central emission.} H$\alpha$ is concentrated chiefly in the central regions of the galaxy. We note that the source of the gas ionisation (AGN, LI(N)ER, or star formation) has yet to be investigated.

\item \textbf{Prominent ring.} The galaxy possesses a prominent ring of H$\alpha$ emission, generally coincident with a ring in the optical image, and frequently with H$\alpha$ emission in the central region. 

\item \textbf{Ends of the bar, or centre and ends.} There is significant H$\alpha$ emission at the ends of the bar, and, more commonly, the central region and the ends of the bar.
 
\item \textbf{No H$\alpha$ detected.} No significant H$\alpha$ emission is detected in this galaxy.

\end{itemize}

\begin{figure}[b]
\begin{center}
 \includegraphics[width=0.9\textwidth]{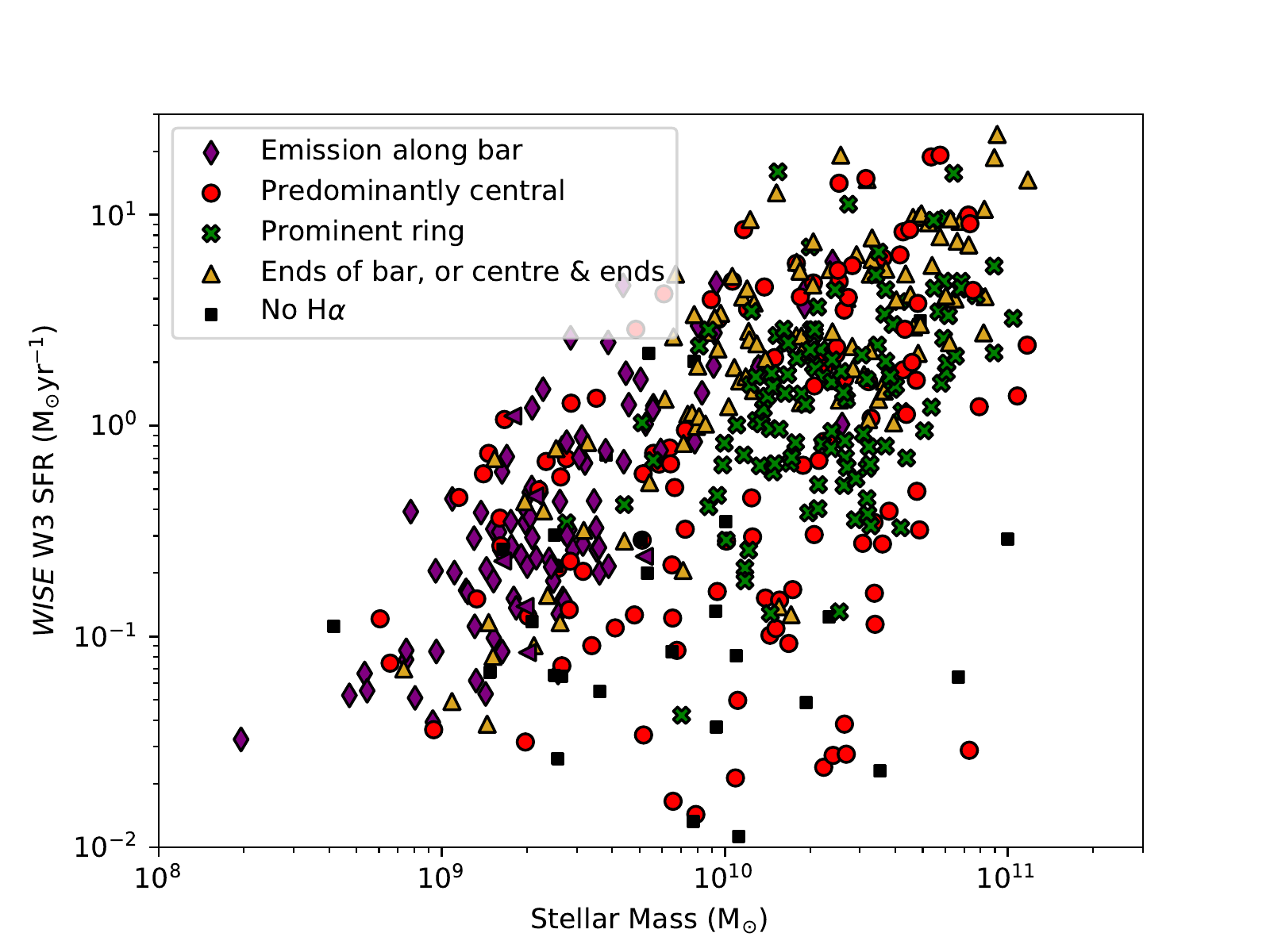} 
 \caption{Star formation main sequence of MaNGA barred galaxies. points are coloured by their H$\alpha$ morphology. It is clear that various H$\alpha$ morphologies inhabit different regions of this diagram.}
   \label{fig3}
\end{center}
\end{figure}

In Fig~\ref{fig3} we again present a SFMS plot, this time with points colour-coded by the morphology of the H$\alpha$ gas in the galaxy. Interestingly, distinct regions are present for given H$\alpha$ morphologies. Galaxies with gas along the bar are low-mass, and lie along the SFMS. Galaxies that exhibit a prominent ring of H$\alpha$ however, are higher mass and usually star forming. H$\alpha$ in the centre and ends of the bar are also found mostly in high-mass galaxies, and central H$\alpha$ is found across the SFMS parameter space.

\section{Discussion \& Conclusions}

 Fig~\ref{fig1} shows that low-mass galaxies tend to host physically shorter bars compared to high-mass galaxies. There may be many reasons for this. Given low-mass galaxies are likely to be gas-rich \cite[(Catinella et al. 2010)]{Catinella10}, and the simulations of \cite[Athanassoula et al. (2013)]{Athanassoula13} have shown that it is far more difficult to form a bar in a gas-rich disk than a gas-poor one, it is possible that these bars have only recently formed, hence their small physical size. If we assume that bars are long-lived structures (which is far from certain), we might imagine a scenario in which bars formed first in the gas-poor disks of high-mass galaxies, and had time to grow to large physical size.  Gas-rich disks, however, took longer to become bar-unstable, and bars present in these disks have formed only recently. 
 Given that only low-mass galaxies host H$\alpha$ (and, presumably, current star formation) along their bars, it is possible that this gas inflow phase is short-lived, and we are witnessing it now for the recently-formed bars embedded within low-mass galaxies. Another possible explanation is that only in low-mass galaxies is the shear low enough along the bar that star formation can take place.
 
 For high-mass galaxies, the story seems to be different. Many high-mass galaxies host rings of H$\alpha$ (accompanied by a visible stellar ring in the optical images). Rings and bar resonances are intimately linked, and it may be that when the bar formed in these galaxies, so too did the strong resonance feature of the inner ring. The ring may halt inflowing ionised gas and stop it from flowing along the bar. 
  A suite of simulations of gas-rich and gas-poor low- and high-mass galaxies are required to test these possibilities.
   
 
In summary, we find that the while physical bar length correlates chiefly with galaxy stellar mass, scaled bar length correlates with galaxy distance below the SFMS. Galaxies that are more passive host bars of longer scaled length (which are presumably more evolved), suggesting a bar quenching picture. When the morphology of the ionised gas is considered, H$\alpha$ emission is only found along the bars of low-mass galaxies. High-mass galaxies are more likely to host H$\alpha$ in a ring-like structure, probably linked to the bar resonances. We conclude that the stellar mass of the galaxy is the most important influence on the bar dynamics and the effect the bar is having on its host galaxy.

\end{document}